\documentstyle[epsfig]{mn2e}
\begin{document}
\input BoxedEPS.tex
\newcommand{\aip}{{\small ${\cal AIPS}$}}
\newcommand{\gtsim}{\mbox{{\raisebox{-0.4ex}{$\stackrel{>}{{\scriptstyle\sim}}
$}}}}
\newcommand{\ltsim}{\mbox{{\raisebox{-0.4ex}{$\stackrel{<}{{\scriptstyle\sim}}
$}}}}
\newcommand{\s}{$\stackrel{\rm s}{.}$}
\newcommand{\h}{$^{\rm h}$}
\newcommand{\m}{$^{\rm m}$}
\newcommand{\pp}{$\stackrel{\prime\prime}{.}$}
\newcommand{\de}{$^{\circ}$}
\newcommand{\p}{$^{\prime}$}
\newcommand{\arc}{$^{\prime\prime}$}
\newcommand{\marc}{^{\prime\prime}}
\newcommand{\rs}{{\em $r_s$}}
\newcommand{\DPM}{{\em DPM}}
\newcommand{\alf}{{\displaystyle\biggl({\nu_{\rm h} \over \nu_{\rm l}}\biggr)^{\alpha}} }

\newcommand{\figstart}[1]
    { \begin{figure}[htb]
      \begin{picture}(0,#1) }
\newcommand{\figend}[4]
    { \end{picture}
      \special{#1}
      \caption[#2]{#3}
      \label{#4}
      \end{figure} }
\newcommand{\fig}[5]
    { \figstart{#1}
      \figend{#2}{#3}{#4}{#5} }
\newcommand{\bHS}{\beta_{\mbox{\scriptsize HS}}}
\newcommand{\bBF}{\beta_{\mbox{\scriptsize BF}}}
\newcommand{\nT}{\nu_{\mbox{\scriptsize T}}}
\newcommand{\et}{E_{\mbox{\scriptsize T}}}
\newcommand{\nTn}{\nu_{\mbox{\scriptsize Tn}}}
\newcommand{\nTf}{\nu_{\mbox{\scriptsize Tf}}}
\newcommand{\tn}{\tau_{x\mbox{\scriptsize n}}}
\newcommand{\tf}{\tau_{x\mbox{\scriptsize f}}}
\newcommand{\xn}{x_{\mbox{\scriptsize n}}}
\newcommand{\xf}{x_{\mbox{\scriptsize f}}}
\newcommand{\yn}{y_{\mbox{\scriptsize n}}}
\newcommand{\yf}{y_{\mbox{\scriptsize f}}}
\newcommand{\lln}{l_{\mbox{\scriptsize n}}}
\newcommand{\llf}{l_{\mbox{\scriptsize f}}}
\newcommand{\Dn}{f(\Delta_{\mbox{\scriptsize n}})}
\newcommand{\Df}{f(\Delta_{\mbox{\scriptsize f}})}
\newcommand{\B}{\mbox{$B$}}
\newcommand{\Bo}{\mbox{$B$}_{0}}

\SetRokickiEPSFSpecial
\HideDisplacementBoxes


\title[Herschel observations and a model for IRAS 08572+3915]
{{\em Herschel} observations and a model for IRAS 08572+3915: a candidate for
the most luminous infrared galaxy in the local ($z < 0.2$) Universe}
\author[A.~Efstathiou et al.]
{\parbox{\textwidth}{A.~Efstathiou,$^{1}$\thanks{~~~~~~~~~~~~~~~~~~E-mail: \texttt{a.efstathiou@euc.ac.cy}}
C.~Pearson,$^{2,3}$
D.~Farrah,$^{4}$
D.~Rigopoulou,$^{2,5}$
J.~Graci\'a-Carpio,$^{6}$
A.~Verma,$^{5}$
H.W.W.~Spoon,$^{7}$ 
J.~Afonso,$^{8,9}$
J.~Bernard-Salas,$^{3,10}$
D.L.~Clements,$^{11}$
A.~Cooray,$^{12}$ 
D.~Cormier,$^{13}$
M.~Etxaluze,$^{14}$
J.~Fischer,$^{15}$
E.~Gonz\'alez-Alfonso,$^{16}$
P.~Hurley,$^{17}$
V.~Lebouteiller,$^{7,18}$
S.J.~Oliver,$^{17}$
M.~Rowan-Robinson,$^{11}$
E.~Sturm$^{6}$
}\vspace{0.4cm}\\
\parbox{\textwidth}{
$^1$School of Sciences, European University Cyprus, Diogenes Street, Engomi, 1516 Nicosia, Cyprus.\\
$^2$Space Science and Technology Department, CCLRC Rutherford Appleton Laboratory, Chilton, Didcot, Oxfordshire, OX11 0QX, UK\\
$^3$Astrophysics Group, Department of Physics, The Open University, Milton Keynes, MK7 6AA, UK\\
$^4$Department of Physics, Virginia Tech, Blacksburg, VA 24061, USA\\
$^5$Oxford Astrophysics, Denys Wilkinson Building, University of Oxford, Keble Rd, Oxford OX1 3RH\\
$^6$Max-Planck-Institut f\"{u}r extraterrestrische Physik, Postfach 1312, D-85741 Garching, Germany\\
$^7$Cornell University, Astronomy Department, Ithaca, NY 14853, USA\\
$^8$Centro de Astronomia e Astrof\'{\i}sica da Universidade de Lisboa,  Observat\'{o}rio Astron\'{o}mico de Lisboa, Tapada da Ajuda, 1349-018 Lisbon, Portugal\\
$^9$Department of Physics, Faculty of Sciences, University of Lisbon, Campo Grande, 1749-016 Lisbon, Portugal\\
$^{10}$Institut d'Astrophysique Spatiale, Universite Paris-Sud 11, F-91405 Orsay, France\\
$^{11}$Astrophysics Group, Imperial College London, Blackett Laboratory, Prince Consort Road, London SW7 2AZ\\
$^{12}$Department of Physics \& Astronomy, University of California, Irvine, CA 92697, USA\\
$^{13}$
Institut f\"ur theoretische Astrophysik, 
Zentrum f\"ur Astronomie der Universit\"at Heidelberg, 
Albert-Ueberle Str. 2, D-69120 Heidelberg, Germany\\
$^{14}$Departamento de Astrof\'isica, Centro de Astrobiolog\'ia, CSIC-INTA, 
Torrej\'on de Ardoz, 28850 Madrid, Spain\\
$^{15}$Naval Research Laboratory, Remote Sensing Division, 4555 Overlook Ave SW, Washington, DC 20375, USA\\
$^{16}$Universidad de Alcal\'a de Henares, Departamento de F\'{\i}sica
y Matem\'aticas, Campus Universitario, E-28871 Alcal\'a de Henares,
Madrid, Spain\\
$^{17}$Department of Physics \& Astronomy, University of Sussex, Falmer, Brighton BN1 9QH, UK\\
$^{18}$Laboratoire AIM, CEA/DSM-CNRS-Universite Paris Diderot, DAPNIA/Service d'Astrophysique, Saclay, F-91191 Gif-sur-Yvette Cedex, France\\
}}
\maketitle 

\begin{abstract}

 We present {\em Herschel} photometry and spectroscopy, carried out as part of the
 {\em Herschel} ULIRG survey (HERUS), and a model for the infrared to submillimeter
 emission of the ultraluminous infrared galaxy IRAS 08572+3915. This source shows
 one of the deepest known silicate absorption features and no polycyclic aromatic
 hydrocarbon (PAH) emission. The model suggests that this object is powered by an
 active galactic nucleus (AGN) with a fairly smooth torus viewed almost edge-on and
 a very young starburst. According to our model the AGN contributes about 90\% of
 the total luminosity of $1.1 \times 10^{13} L_\odot$, which is about a factor of
 five higher than previous estimates. The large correction of the luminosity is due
 to the anisotropy of the emission of the best fit torus. Similar corrections
 may be necessary for other local and high-z analogs. This correction implies that 
IRAS 08572+3915 at a redshift of 0.05835 may be the nearest hyperluminous infrared
 galaxy and probably the most luminous infrared galaxy in the local ($z < 0.2$)
 universe. IRAS 08572+3915  shows a low ratio of [CII] to IR luminosity (log
 $L_{[CII]}/L_{IR}  < -3.8$) and a [OI]63$\mu m$ to [CII]158$\mu m$ line ratio
 of about 1 that supports the model presented in this letter.
\end{abstract}


\section{Introduction}
\label{}

 Recent results from {\em Herschel} surveys have shown that high-redshift
 ultraluminous infrared galaxies (ULIRGs with luminosities greater than
 $10^{12}L_\odot$) are generally colder and less obscured than their local
 counterparts (Elbaz et al. 2011). This supports pre-{\em Herschel} 
predictions on the basis of models of the spectral energy distributions
 (SEDs) of submillimeter galaxies that they have significant contributions
 to their luminosity from cirrus or cold diffuse dust in the galaxy (Efstathiou
 \& Rowan-Robinson 2003, Efstathiou \& Siebenmorgen 2009). The other
 surprising result of {\em Herschel} and {\em Spitzer} surveys, however,
 is that there are large numbers of distant ULIRGs that show stronger
 silicate absorption than the majority of local ULIRGs (Rowan-Robinson
 et al. 2010, Sajina et al. 2012). 

 IRAS 08572+3915 surely ranks as one of the most peculiar local ULIRGs 
but may be typical of the highly obscured distant ULIRGs. It shows extremely
 deep silicate aborption features both at $9.7$ (silicate strength $\sim -4$;
 see equation 1 of Spoon et al. 2007) and $18\mu m$ (Dudley \& Wynn-Willams
 1997), which are deeper than those of the prototypical ULIRG Arp220, and 
at the same time it shows no evidence of PAH emission features either in the
 mid-infrared or at 3.3$\mu m$ (Imanishi et al. 2008, Veilleux et al. 2009).
 This has been interpreted as evidence that IRAS 08572+3915 is powered by an
 active galactic nucleus (AGN), but results from calculations of the emission
 from clumpy tori, which are favored on theoretical grounds (e.g. Krolik \& 
Begelman 1988), do not show silicate absorption features as deep as those
 observed in IRAS 08572+3915 (Levenson et al. 2007). This is because in a
 clumpy medium, even in the case where we view the torus edge-on, it is
 possible to see the inner hot dust through holes in the cloud distribution
 and this has the effect of filling in the absorption features produced by
 foreground clumps. So for the mid-IR emission from IRAS 08572+3915 to be
 powered by an AGN, the torus must have either an unusually high filling
 factor, or be fairly smooth.

 Spoon et al. (2007) presented a diagnostic diagram that plots the silicate
 strength of about 200 infared galaxies and AGN versus their 6.2$\mu m$ PAH
 equivalent width. Rowan-Robinson \& Efstathiou (2009) showed that the
 distribution of galaxies on the diagram of Spoon et al. can be understood
 in terms of the starburst models of Efstathiou, Rowan-Robinson \& Siebenmorgen
 (2000, hereafter referred to as ERRS00 models) and the AGN torus models of
 Efstathiou \& Rowan-Robinson (1995). The diagonal locus on the diagram can
 be explained by the evolution of the spectral energy distribution (SED) of
 the starbursts in the age range 0-72Myr, and the horizontal locus can be
 explained by mixing of starburst and AGN torus emission. Rowan-Robinson \&
 Efstathiou (2009) suggested that objects such as IRAS 08572+3915, that are
 located at the upper tip of the diagonal locus in class 3A of Spoon et al.,
 and which show very deep silicate absorption and no PAH features, could be
 either AGN with a fairly smooth edge-on torus or very young starbursts. The
 far-infrared to submilimeter colours predicted by these two models are
 however distinctly different so modeling of the complete SEDs with radiative
 transfer models offers the possibility of breaking the degeneracy. 

 This is one of a series of papers that discuss results from an analysis of the
 data collected by the {\em Herschel} ULIRG survey (HERUS; Farrah et al. 2013).
 As part of this survey, 43 ULIRGs have been observed in spectroscopic and
 photometric mode using the SPIRE (Griffin et al. 2010) and PACS (Poglitch et
 al. 2010) instruments onboard {\em Herschel}. In this letter we use HERUS data,
 and data collected as part of the SHINING program (Sturm et al. 2010), as well
 as radiative transfer models for starburst and AGN torus emission to shed light
 on the origin of the luminosity of IRAS 08572+3915. Understanding this object
 is important as it could serve as a template for objects with deep silicate
 absorption that reside at high redshift. In section 2 we describe the {\em
 Herschel} data, in section 3 we describe the models and in section 4 we present
 the best fit model and discuss our results. A flat Universe is assumed with
 $\Lambda =0.73$ and H$_0$=71km/s/Mpc.

\section{Data}

IRAS 08572+3915 at a redshift of 0.05835 was observed by the {\it Herschel Space
 Observatory}  (Pilbratt et al. 2010) on 10th October 2011 with the SPIRE Photometer
 (Griffin et al. 2010). The photometer observations were made in Small Map Mode
 in the PSW (250$\mu$m), PMW (350$\mu$m) \& PLW (500$\mu$m) bands. The observation
 was processed using Version 8 of the Herschel Common Science System  {\it Herschel
 Interactive Processing Environment} (HIPE, Ott et al. 2010) using the standard
 user pipeline (Dowell et al. 2010), with default values for all tasks utilizing
 the SPIRE Calibration Tree version 8.1. Standard median baseline removal was made
 to create the final images using the naive mapper task. Photometry of the source
 was made using the SPIRE Timeline Source Fitter Task in HIPE which fits a 2-D
 Gaussian to the timeline data at the coordinates of the source. We assumed  FWHM
 of 18.15, 25.2, 36.9 arcsec for the PSW, PMW, PLW bands respectively. The background
 was measured within an annulus between 300 to 350 arcsec and then an elliptical
 Gaussian function was fit to both the central 22, 32, 40 arcsec (for the PSW, 
PMW, PLW bands respectively) and the background annulus. The output is the fitted
 flux, RA, Dec and associated errors. The fitted fluxes obtained from the source
 fitter are given in Table 1.

IRAS 08572+3915 was observed  with the SPIRE Fourier Transform Spectrometer
 (Griffin et al. 2010) on the  7th November 2011. The observation was made
 in High Resolution Point Source Mode covering the entire submillimeter spectrum
 in two detector arrays from 194-313$\mu$m (SSW) and 303-671$\mu$m (SLW). The
 spectrometer observations were also reduced with the standard HIPE version 11
 spectrometer user pipeline  (Fulton et al. 2010), processing just the central
 detectors of each array to produce the final calibrated point source spectra.
 The standard pipeline subtracts the telescope background emission using an 
emission model derived from the measured primary and secondary mirror temperatures
 during the observation. This process leaves an uncertainty of $\sim$1.5Jy, 
 significant for faint sources. A more accurate background subtraction was
 therefore made using the off-axis detectors in each array to measure areas
 of dark sky. The resulting spectrum for IRAS 08572+3915 shows three $^{12}$CO
 lines ($^{12}$CO(11-10)~237$\mu$m, $^{12}$CO(10-9)~260$\mu$m and $^{12}$CO(9-8)~289$\mu$m)
  and the $[CI]$ line at a rest wavelength of 607$\mu$m. More details about
 these results will be given in a forthcoming paper.

We also obtained PACS spectroscopy of IRAS 08572+3915 as part of the SHINING
 program (PI E.Sturm). The data reduction of the PACS spectroscopic observations
 was carried out using the standard {\em Herschel} data reduction pipeline
 included in Hipe 6.0, with some modifications introduced to correct for
 small offsets in the continum of the spectral pixels. The spectrum was 
normalized to the telescope background and recalibrated with a reference Neptune
 spectrum obtained during the Herschel PV phase. With this method errors
 in the absolute flux calibration are generally lower than 20\%. The galaxy
 is detected in three far-infrared lines whose fluxes are given in Table 1.

 IRAS 08572+3915 was one of the ULIRGs observed in the Spitzer Guaranteed
 Time Observation (GTO) ULIRG project (PID 105; P.I.: J. R. Houck) and its 
IRS spectrum first appeared in Spoon et al. (2006). In the analysis that
 follows we will use the IRS spectrum resulting from the data processing
 of Wang et al. (2011). Similar results are obtained by using the spectrum
 available by the CASSIS database (Lebouteiller et al. 2011).  

\begin{table}
\caption{SPIRE photometry and PACS spectroscopy for IRAS 08572+3915.}
\begin{center}
\begin{tabular}{l l l}
\hline
\hline
Photometry  & & \\
\hline
Wavelength  &  Flux (Jy)& Error (mJy)  \\
\hline
250$\mu$m   & 0.532	 &  4.1 \\
350$\mu$m   & 0.168	 &  4.0 \\
500$\mu$m   & 0.056  &   4.8 \\
      &         &         \\
Spectroscopy  & &\\
\hline
Wavelength  &  Intensity (W m$^{-2}$)& Error (W m$^{-2}$)  \\
\hline
\\$[OI]$~63$\mu$m & $1.18 \times 10^{-16}$  & $1.74 \times 10^{-17}$  \\
$[OIII]$~88$\mu$m & $5.73 \times 10^{-17}$  & $2.63 \times 10^{-18}$ \\		   	
$[CII]$~158$\mu$m & $1.12 \times 10^{-16}$  & $3.00 \times 10^{-18}$  \\

\hline
\end{tabular}
\end{center}
\end{table}

\section{Radiative transfer models}

Several models for the infrared emission of starburst galaxies have been developed
 (Rowan-Robinson \& Crawford 1989, Rowan-Robinson \& Efstathiou
 1993, Kr\"ugel \& Siebenmorgen 1994, Silva et al. 1998, Takagi
 et al. 2003, Dopita et al. 2005, Siebenmorgen \& Kr\"ugel 2007).
 Efstathiou, Rowan-Robinson \& Siebenmorgen (2000) presented a
 starburst model that combined the stellar population synthesis
 model of Bruzual \& Charlot, radiative transfer
that included the effect of small grains and PAHs, and a simple
 scheme for the evolution of the molecular clouds that constitute
 the starburst. The model predicts the spectral energy distributions
 of starburst galaxies from the ultraviolet to the millimetre 
 for different ages of the starburst and different initial optical depths
 of the molecular clouds. The ERRS00 model uniquely predicts that young starbursts have small PAH
 equivalent widths and deep silicate absorption features whereas older
 starbursts have stronger PAH features and more shallow silicate absorption
 features. In this paper we use a grid of starburst models which have been
 computed with the method of ERRS00 but with a revised dust model
 (Efstathiou \& Siebenmorgen 2009). In this grid of models we vary
the initial optical depth in the V band of the molecular clouds ($\tau_V=50$,
 75,  and 100) and the age of the starburst in the range 0-70Myrs in steps of 5Myr.

Radiative transfer models of the torus in AGN have been presented by Pier  \& Krolik
 (1992), Granato \& Danese (1994), Nenkova et al. (2002, 2008), Dullemond \& van Bemmel
 (2005), H\"onig et al. (2006), Schartmann et al. (2008), Stalevski et al. (2012), 
Heymann \& Siebenmorgen (2012). Efstathiou \& Rowan-Robinson (1995) considered three
 different types of geometry for the torus and concluded that the geometry that best
 fitted the observational constraints was that of tapered discs (whose thickness
 increases linearly with distance from the central source in the inner part of the
 disc but tapers off to a constant height in the outer part). The tapered disc models
 consider a distribution of grain species and sizes, multiple scattering and a smooth
 density distribution that follows $r^{-1}$ where $r$ is the distance from the central
 source. The models  have been quite successful in fitting the spectral energy
 distributions of AGN even in cases where mid-infrared spectroscopy is available
 (Efstathiou et al. 1995, Alexander et al. 1999, Ruiz et al. 2001, Farrah et al. 2002,
 Verma et al. 2002, Farrah et al. 2003, Efstathiou \& Siebenmorgen 2005, Farrah et al. 
2012, Efstathiou et al. 2013).

\begin{figure}
\epsfig{file=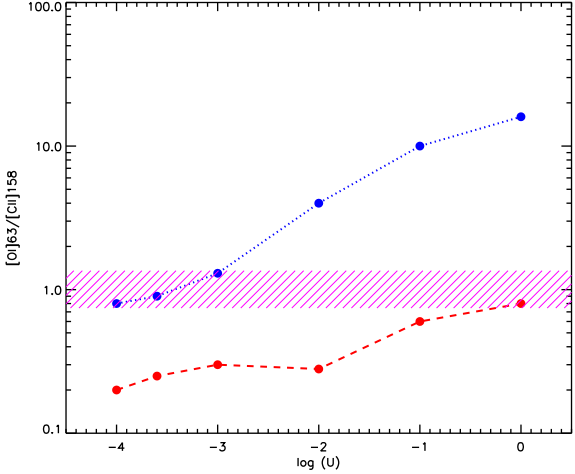, angle=0, width=8cm}
\caption{
Line ratios predicted from CLOUDY runs with different ionization
parameters $U$. The dotted blue line corresponds to an AGN spectrum
whereas the dashed red line corresponds to a zero-age starburst computed
with the Bruzual \& Charlot (2003) models and a Salpeter IMF. The magenta shaded
 region indicates the measured [OI]63$\mu m$/[CII]158$\mu m$ ratio and its
 2$\sigma $ uncertainty. 
}
\end{figure}
   
In this paper we use a grid of tapered disc models computed with the method of
 Efstathiou \& Rowan-Robinson (1995) and described in more detail in Efstathiou
 et al. (2013). In this grid of models we consider five discrete values for the
 equatorial 1000\AA~ optical depth ($\tau_{UV}^{eq}=$250, 500, 750, 1000, 1250;
 $\tau_{UV}^{eq} \approx 5 \tau_{V}^{eq}$), three  values for the ratio of outer
 to inner disc radii ($r_2/r_1=$20, 60, 100) and four values for the half-opening
 angle of the disc ($\Theta_0=$30$^o$, 45$^o$, 60$^o$ and 75$^o$; $\Theta_1$ as
 defined by Efstathiou \& Rowan-Robinson is equal to $90-\Theta_0$).  The spectra
 are computed for 37 inclinations $i$ which are equally spaced in the range 0 to $90^o$.   
The grids of tapered disc and starburst models discussed above are available from
 AE on request (a.efstathiou@euc.ac.cy). 

\begin{figure}
\epsfig{file=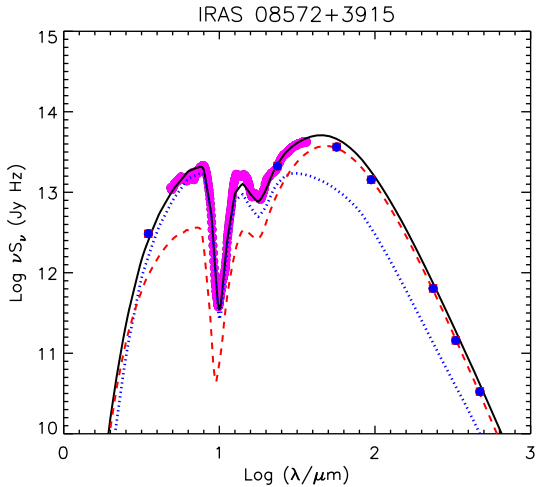, angle=0, width=8cm}
\caption{
Fit to the spectral energy distribution of IRAS 08572+3915 with a smooth edge-on
 torus (blue dotted line) and a young starburst (red dashed line). The total
 emission is given by the black solid line. Broad band data (blue filled circles)
 are from this work, Carico et al. (1988) and  IRAS. The Spitzer/IRS spectrum 
(Wang et al. 2011) is shown in magenta.
}
\end{figure}

\begin{figure}
\epsfig{file=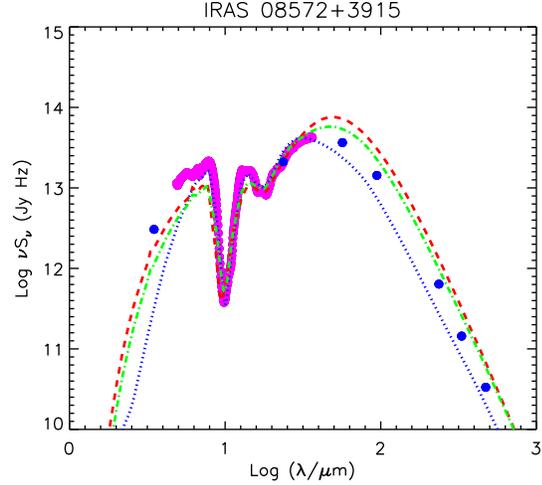, angle=0, width=8cm}
\caption{
Best fit models to the spectral energy distribution of IRAS 08572+3915 with pure
 smooth torus (blue dotted line), pure clumpy torus (green dot dashed line) and
 pure starburst (red dashed line). The data are the same as those plotted in Figure 2.
}
\end{figure}   

\section{Results and discussion}

 In Figure 1 we compare the observed [OI]63$\mu m$ to [CII]158$\mu m$
 line ratio with photoionization models computed with CLOUDY (Ferland et al. 1998). We have
 computed models that assume different ionization parameters $U$ and
 two different source spectra: an AGN spectrum (shown with the blue
 dashed line in Figure 1) and a young starburst spectrum that assumes
 a zero-age burst of star formation (red dotted line). The young starburst
 spectrum, which is identical with the source spectrum that is assumed by
 the dust radiative transfer model (see next paragraph), is obtained from
 the tables of Bruzual \& Charlot (2003) for a Salpeter IMF and solar metallicity. 
In all CLOUDY runs we assume a density at the illuminated face of the cloud of
 $10^3$ cm$^{-3}$ and a total column density of $2 \times 10^{23}$ cm$^{-2}$.

It is clear from Figure 1 that the observed line ratio is consistent with either
 an AGN spectrum with a low ionization parameter or a young starburst spectrum
 with a high ionization parameter. Another indicator of activity in galaxies is
 the ratio of [CII] to IR luminosity. IRAS 08572+3915  shows a low ratio (log 
$L_{[CII]}/L_{IR}$=-3.8 assuming conservatively that $L_{IR} = 1.5 \times 10^{12}
 L_\odot$) which places it well in the range shown by systems dominated by an AGN
 (Abel et al. 2009, Sargsyan et al. 2012). Detailed modelling of the PACS and SPIRE
 line observations will be presented in a future paper.

 We can now combine the submillimeter photometry presented here with the Spitzer
 IRS spectrum from 5-35$\mu m$, IRAS data and near-IR photometry to get another
 constraint on the power source. In Figure 2 we present our best fit model for 
the spectral energy distribution of IRAS 08572+3915 which is obtained by finding
 the combination of starburst and smooth AGN torus models that minimizes $\chi^2$.
 Our fitting code can return a pure AGN or a pure starburst model if it is the
 preferred solution. We find that a pure AGN model (shown in Figure 3) has a 
reduced $\chi^2_{min}$ of 9.1 whereas the corresponding value for a pure starburst
 model is 15.8. The $\chi^2_{min}$ of the AGN/starburst combination is 3.25. The
 best fit starburst model has an age of 0Myr and an initial optical depth of the
 molecular clouds that constitute the starburst of 75. The best fit torus model
 assumes $r_2/r_1=60$, $\tau_{UV}^{eq}=500$, $\Theta_0=75^o$ and $i=88^o$. Our
 grid of starburst and AGN torus models is fairly crude so we can get sensible
 uncertainties only on the inclination and age of the starburst. We find that
 even for $\Delta \chi^2 =30$ the uncertainty in inclination is $5^o$ and in
 age is 5Myr. The 1-1000$\mu m$ starburst luminosity is predicted to be $10^{12}
 L_\odot$ and the AGN {\em apparent} luminosity $6.6 \times 10^{11} L_\odot$.
 Because of the anisotropy of the emission of the  torus, the luminosity of 
the AGN must be multiplied by the anisotropy correction factor $A$ of 14.6
 to give the intrinsic luminosity of $9.6 \times 10^{12} L_\odot$ (Efstathiou 
2006). Assuming there is no processing of the AGN and starburst emission
 by the host galaxy, the total luminosity of the system is therefore predicted
 to be $1.1 \times 10^{13} L_\odot$, 90\% of which is due to the AGN.  According
 to Rowan-Robinson \& Wang (2010; see their Figure 1) the nearest IRAS galaxy
 that exceeds the $10^{13} L_\odot$ threshold, and is therefore classified as
 a hyperluminous infrared galaxy, lies at $z > 0.3$.
 Other IRAS galaxies within $z < 0.3$ may need a correction of their luminosity
 because of anisotropic torus emission but such high anisotropy corrections are only
 expected for AGN with very deep silicate absorption features. IRAS 08572+3915
 has the second deepest silicate feature in the sample of Spoon et al. (2007), 
 just shallower than that
 of IRAS 01298-0744 at a redshift of 0.13618 which is also the galaxy with the
 highest silicate optical depth in the sample of Veilleux et al. (2009). Although
 according to our model  IRAS 01298-0744 also needs a large anisotropy correction,
 its intrinsic luminosity is estimated to be only about $5 \times 10^{12}  L_\odot$.
 IRAS 08572+3915 may therefore be the nearest hyperluminous infrared galaxy and
 the most luminous infrared galaxy in the local ($z < 0.2$) Universe. The solution
 found here is different from that found by Farrah et al (2003) who only considered
 broad band data and estimated that the starburst in IRAS 08572+3915 is about a
 factor of two more luminous than the AGN and the galaxy has a total IR luminosity
 of $1.5 \times 10^{12} L_\odot$.

To test whether clumpy torus models can match the spectral energy distribution
 of IRAS 08572+3915, we have compared its SED with the models of Stalevski et al.
 (2012). We find that even the AGN torus models with the most optically thick
 clumps ($\tau_{9.7\mu m}=10$) when observed edge-on fail to match the deep
  silicate absorption feature of IRAS 08572+3915 ($\chi^2_{min}=28.8$; see
 Figure 3). A combined clumpy torus and starburst fit gives a $\chi^2_{min}$ of 13.3.

Determining the true distribution of dust in the torus (i.e. whether it is smooth,
 filamentary, clumpy or two-phase medium) is important for understanding the physics
 of the torus but also for estimating the intrinsic luminosity of AGNs from the
 observed one. Clumpy tori generally emit much more isotropically than smooth tori
 (Nenkova et al 2008, H\"onig et al 2011). H\"onig et al (2011) estimate that at
 15$\mu m$ type 1 AGN are only a factor of 1.4 more luminous than type 2 AGN but
 according to the best fit torus model presented here for IRAS 08572+3915, at
 15$\mu m$ the ratio of face-on/edge-on emission is of order 10 and this is what
 makes this and other similar objects so intrinsically luminous.

\section*{Acknowledgments}

{\em Herschel} is an ESA space observatory with science instruments provided
 by European-led Principal Investigator consortia and with important participation
 from NASA. Basic research in IR astronomy at NRL is funded by the US ONR. 
 JF also acknowledges support from the NHSC. JBS acknowledges support from
 a Marie Curie Intra-European Fellowship within the 7th European Community
 Framework Program under project number 272820. 
E.G-A is a Research Associate at the Harvard-Smithsonian
Center for Astrophysics, and thanks the Spanish
Ministerio de Econom\'{\i}a y Competitividad for support under projects
AYA2010-21697-C05-0 and FIS2012-39162-C06-01. 
ME acknowledges support from ASTROMADRID (grant S2009ESP-1496), ASTROMOL
 (CSD2009-00038) and the Spanish MINECO (AYA2009-07304, and AYA2012-32032).
 JA acknowledges support from the Science and Technology Foundation
 (FCT, Portugal) through the research grants PTDC/CTE-AST/105287/2008, 
 PEst-OE/FIS/UI2751/2011 and PTDC/FIS-AST/2194/2012.  We also thank
 an anonymous referee for his comments and suggestions.

\end{document}